\PassOptionsToPackage{hidelinks,breaklinks=true}{hyperref}
\PassOptionsToPackage{inline}{enumitem}
\documentclass[wcp]{jmlr}
\usepackage{longtable}

\usepackage{booktabs}
\usepackage{pifont} 
\makeatletter
\expandafter\let\csname enumerate*\endcsname\relax
\expandafter\let\csname endenumerate*\endcsname\relax
\expandafter\let\csname itemize*\endcsname\relax
\expandafter\let\csname enditemize*\endcsname\relax
\expandafter\let\csname description*\endcsname\relax
\expandafter\let\csname enddescription*\endcsname\relax
\makeatother
\usepackage{enumitem} 
\usepackage{algorithm}
\usepackage[noend]{algorithmic}
\usepackage[utf8]{inputenc}
\usepackage{amsmath}
\usepackage{amsfonts}
\usepackage{mathtools}
\usepackage{booktabs} 
\usepackage{comment}
\usepackage{amsmath,graphicx}
\usepackage{multirow}
\usepackage{amssymb}
\usepackage{mathtools}
\usepackage{breakurl}
\usepackage{epstopdf}
\usepackage{lipsum}
\usepackage{soul}
\usepackage{times}
\usepackage{makecell}
\usepackage{color,soul}
\soulregister\cite7
\soulregister\ref7
\soulregister\pageref7
\renewcommand{\footnotesize}{\scriptsize}
\usepackage{wrapfig}
\usepackage{soul}
\usepackage{comment}
\usepackage{caption}
\usepackage{lipsum}
\usepackage{microtype}
\newcommand\eat[1]{} 
\usepackage{footnote}
\makesavenoteenv{tabular}
\usepackage{setspace}
\usepackage{titlesec}

\usepackage{subcaption} 
\usepackage{graphicx}
\usepackage{subcaption}

\usepackage{multirow}
\usepackage{soul}
\usepackage{xcolor}
\colorlet{hidden-draw}{gray!60}   
\colorlet{hidden-pink}{pink}      
\colorlet{darkgray}{gray!30}

\sethlcolor{green!40}
\newcommand{\hlred}[1]{{\sethlcolor{red!20}\hl{#1}}}

\usepackage{caption}
\usepackage{longtable} %

\usepackage{soul}
\usepackage[normalem]{ulem}
\usepackage{bm}

\newtheorem{mydef}{Definition}
\usepackage{wrapfig}

\usepackage[edges]{forest}
\usepackage{tikz}
\usetikzlibrary{shapes, arrows.meta, positioning}
\titlespacing*{\subsection}{0pt}{0.5\baselineskip}{0.3\baselineskip}

\tikzstyle{my-box}=[rectangle, draw=hidden-draw, rounded corners, align=left, text opacity=1, minimum height=1.5em, minimum width=7em, inner sep=2pt, fill opacity=.8, line width=0.8pt]
\tikzstyle{leaf-head}=[my-box, draw=gray!80, fill=gray!15, text=black, font=\normalsize, inner xsep=2pt, inner ysep=4pt]
\tikzstyle{red-box}=[my-box, draw=red!70, fill=red!15, text=black, font=\normalsize, inner xsep=2pt, inner ysep=4pt]
\tikzstyle{leaf-red-box}=[my-box, draw=red!80, fill=hidden-pink!30, text=black, font=\normalsize, inner xsep=3pt, inner ysep=6pt]

\tikzstyle{blue-box-lv2}=[my-box, draw=cyan!70, fill=cyan!15, text=black, font=\normalsize, inner xsep=2pt, inner ysep=4pt]
\tikzstyle{blue-box-lv3}=[my-box, draw=cyan!70, fill=cyan!15, text=black, font=\normalsize, inner xsep=2pt, inner ysep=4pt]
\tikzstyle{blue-box-lv4}=[my-box, draw=cyan!100, fill=white, text=black, font=\normalsize, inner xsep=2pt, inner ysep=4pt]

\tikzstyle{green-box-lv2}=[my-box, draw=green!70, fill=green!15, text=black, font=\normalsize, inner xsep=2pt, inner ysep=4pt]
\tikzstyle{green-box-lv3}=[my-box, draw=green!70, fill=hidden-pink!11, text=black, font=\normalsize, inner xsep=2pt, inner ysep=4pt]
\tikzstyle{green-box-lv4}=[my-box, draw=green!100, fill=white, text=black, font=\normalsize, inner xsep=2pt, inner ysep=4pt]

\tikzstyle{orange-box-lv2}=[my-box, draw=orange!70, fill=orange!15, text=black, font=\normalsize, inner xsep=2pt, inner ysep=4pt]
\tikzstyle{orange-box-lv3}=[my-box, draw=orange!70, fill=hidden-pink!11, text=black, font=\normalsize, inner xsep=2pt, inner ysep=4pt]
\tikzstyle{orange-box-lv4}=[my-box, draw=orange!100, fill=white, text=black, font=\normalsize, inner xsep=2pt, inner ysep=4pt]

\tikzstyle{violet-box-lv2}=[my-box, draw=violet!70, fill=violet!15, text=black, font=\normalsize, inner xsep=2pt, inner ysep=4pt]
\tikzstyle{violet-box-lv3}=[my-box, draw=violet!70, fill=hidden-pink!11, text=black, font=\normalsize, inner xsep=2pt, inner ysep=4pt]
\tikzstyle{violet-box-lv4}=[my-box, draw=violet!100, fill=white, text=black, font=\normalsize, inner xsep=2pt, inner ysep=4pt]

\usepackage{lineno}

\makeatletter
\let\Ginclude@graphics\@org@Ginclude@graphics 
\let \@jmlrpages \@empty
\makeatother

\jmlrvolume{304,}
\jmlryear{2025}
\jmlrworkshop{ACML 2025}

\title[$\delta$-\textsc{Steal}:  LLM Stealing Attack with LDP]{$\delta$-\textsc{Steal}:  LLM Stealing Attack  with Local Differential Privacy}

\author{\Name{Kieu Dang}$^{\star}$ \Email{vdang@albany.edu}\\
\Name{Phung Lai}$^{\star}$ \Email{lai@albany.edu}\\
\Name{NhatHai Phan}$^{\ddagger}$ \Email{phan@njit.edu}\\
\Name{Yelong Shen}$^{\S}$ \Email{Yelong.Shen@microsoft.com}\\
\Name{Ruoming Jin}$^{\P}$ \Email{rjin1@kent.edu}\\
\Name{Abdallah Khreishah}$^{\ddagger}$ \Email{abdallah@njit.edu}\\
\addr $^{\star}$University at Albany, 
$^{\ddagger}$New Jersey Institute of Technology, 
$^{\S}$Microsoft, 
$^{\P}$Kent State University
}
\editors{Hung-yi Lee and Tongliang Liu}

\begin{document}

\maketitle

\begin{abstract}
Large language models (LLMs) demonstrate remarkable capabilities across various tasks. However, their deployment introduces significant risks related to intellectual property. In this context, we focus on model stealing attacks, where adversaries replicate the behaviors of these models to steal services. These attacks are highly relevant to proprietary LLMs and pose serious threats to revenue and financial stability. To mitigate these risks, the watermarking solution embeds imperceptible patterns in LLM outputs, enabling model traceability and intellectual property verification.

In this paper, we study the vulnerability of LLM service providers by introducing $\delta$-\textsc{Steal}, a novel model stealing attack that bypasses the service provider's watermark detectors while preserving the adversary's model utility.  $\delta$-\textsc{Steal} injects noise into the token embeddings of the adversary's model during fine-tuning in a way that satisfies local differential privacy (LDP) guarantees. The adversary queries the service provider's model to collect outputs and form input-output training pairs. By applying LDP-preserving noise to these pairs, $\delta$-\textsc{Steal} obfuscates watermark signals, making it difficult for the service provider to determine whether its outputs were used, thereby preventing claims of model theft. Our experiments show that $\delta$-\textsc{Steal} with lightweight modifications achieves attack success rates of up to $96.95\%$ without  significantly compromising the adversary's model utility. The noise scale in LDP controls the trade-off between attack effectiveness and model utility. This poses a significant risk, as even robust watermarks can be bypassed, allowing adversaries to deceive watermark detectors and undermine current intellectual property protection methods. 
  
\end{abstract}
\begin{keywords}
LLMs; Stealing Attacks; Watermarks; Local Differential Privacy
\end{keywords}

\section{Introduction}

Large language models (LLMs), such as ChatGPT \citep{OpenAI} or Gemini \citep{gemini}, have demonstrated remarkable capabilities in diverse tasks such as text generation, machine translation, and reasoning \citep{li2024pre}. Despite their widespread adoption, training these models demands significant computational resources and human effort, leading to their common deployment as services through APIs \citep{minthigpen}. 
While API access restricts direct access to model weights, it does not prevent adversaries from exploiting the model's outputs, undermining the intellectual property protection of proprietary models. This prevents a service provider from protecting ownership over its generated content.
Adversaries can exploit generated outputs to fine-tune their local models, effectively replicating the LLM's behavior in specific domains if collecting sufficient data \citep{carlini2024stealing, jovanovic2024watermark}.  
These deployment risks highlight significant concerns about the intellectual property protection of the service provider's proprietary LLMs.  

To mitigate risks, service providers have implemented various strategies, including  watermarking, encryption, and restricted  API access \citep{kirchenbauer2023watermark,xue2022advparams}. Among them, watermarking techniques   \citep{christ2024undetectable,kirchenbauer2023watermark,kuditipudi2023robust} are especially effective, offering traceability and scalability. Typically,  watermarks inject imperceptible patterns into LLM's outputs. Thanks to these unique patterns, once a watermark is applied, the service provider can use watermark detectors to verify the existence of watermarks in generated text, enabling traceability and detection of intellectual property violations through content tracking.

To bypass watermark detectors and replicate the behavior of LLMs, adversaries  employ various techniques, including watermark removal  \citep{krishna2024paraphrasing,pan2024markllm,zhang2023watermarks,kirchenbauer2023watermark} and model stealing \citep{jovanovic2024watermark,carlini2024stealing,wu2024bypassing,birch2023model}. Watermark removal typically involves modifying outputs through synonym substitution, paraphrasing, or sentence restructuring to erase traceable patterns. Model stealing, on the other hand, attempts to extract watermark components from service providers via API queries. However, these attacks face key challenges as they often degrade model utility by increasing perplexity or distorting the semantic meaning of watermarked outputs \citep{ren2023robust,zhang2023watermarks}, and they are typically limited to specific watermark types, recovering only small portions of the original model. As a result, it becomes difficult for adversaries to replicate the behaviors of LLMs without impacting their functionality or output quality.

\textbf{Key contributions.} 
To balance the trade-off between attack effectiveness and model utility, we introduce a novel model stealing attack, called $\delta$-\textsc{Steal},  designed to bypass watermark detectors while maintaining high model utility. 
The key idea is to leverage the concept of local differential privacy (LDP) to obscure the differences between watermarked and non-watermarked LLM outputs.
By injecting LDP-preserving noise into the token embeddings of the adversary's model during fine-tuning, our attack makes watermarked outputs indistinguishable from non-watermarked ones, hindering accurate detection. This prevents service providers from verifying model ownership and protects the stolen model from being detected. Our key contributions are summarized as follows.
\begin{itemize}
    \item[$\bullet$]  \textbf{$\delta$-\textsc{Steal}}: We propose  $\delta$-\textsc{Steal} LLM stealing attack, which incorporates LDP techniques to ensure that   the adversary's model retains its effectiveness, functionality, and  semantic quality of its outputs, closely resembling that of the service provider's model.
    
\item[$\bullet$] \textbf{Controlling the Trade-off between Attack Effectiveness and Model Utility}: Adjusting the noise scale $\delta$, we can  balance the trade-off between attack effectiveness and  utility, offering a promising solution for adversaries to steal model behaviors without sacrificing utility.
    
\item[$\bullet$] \textbf{Generality and Flexibility}: 
    $\delta$-\textsc{Steal} is model-agnostic and watermark-agnostic, allowing it to function across different LLMs and watermarks. This versatility makes it highly adaptable and practically applicable in a variety of real-world scenarios.
   \item[$\bullet$] \textbf{Experimental Validation}: 
   Our experiments demonstrate that through lightweight modifications during the fine-tuning process of the adversary's local LLM, $\delta$-\textsc{Steal} effectively bypasses existing watermarks across different LLMs and attacks, achieving attack success rates of up to 96.95\% while being supported by theoretical guarantees. 
   
\end{itemize}

Code and appendices are at: {\small \url{https://github.com/kirudang/LDP_Stealing_Attack}}

 \begin{table*}[t!]
    \caption{A summary of $\delta$-\textsc{Steal} and related work. WM stands for watermark.}
    \renewcommand{\arraystretch}{1.2}
    \footnotesize
    \begin{tabular}{lccccccc}  
        \toprule
        \makecell{Attack} & \makecell{Objectives} & \makecell{Scope} & \makecell{Guarantee} & \makecell{Modifications}  \\
        \hline
        \hline
\makecell{Dipper \\ \citep{krishna2024paraphrasing}} & \makecell{Sentence-level \\ WM removal}  & \makecell{WM-agnostic} &  \ding{55} & \makecell{Modify WM sentences \\ using its  pre-trained model} \\
\hline
\makecell{Substitution \\ \citep{pan2024markllm}} & \makecell{Token-level \\  WM removal} & \makecell{WM-agnostic} &  \ding{55} & \makecell{Modify WM text \\with synonyms} \\
\hline
\makecell{WMRemoval \\ \citep{zhang2023watermarks}} & \makecell{Token-level \\ WM removal} & \makecell{WM-agnostic}  &   \ding{55} &\makecell{Modify WM text \\with quality check} \\
\hline
\makecell{CoRPG \\ \citep{lin2021towards}} & \makecell{Paragraph-level \\ WM removal} & \makecell{WM-agnostic} &  \ding{55}& \makecell{Modify WM text  based on \\sentence relationship} \\
\hline
\makecell{Spoofing \\ \citep{jovanovic2024watermark}} & \makecell{Steal WM rules}  & \makecell{Green-red WM (KGW  family \\ \citep{kirchenbauer2023watermark})}  &  \ding{55} & \makecell{Learn WM rules\\ from target LLM's output}\\
\hline
\makecell{Color-Substitution \\ \citep{wu2024bypassing}} & \makecell{Steal WM \\ vocabulary lists}  & \makecell{Green-red WM (KGW  family \\ \citep{kirchenbauer2023watermark})}&  \ding{51} & \makecell{Observe LLM's outputs \\ by repeating API queries}\\
\hline
\makecell{Part Stealing\\ \citep{carlini2024stealing}} & \makecell{Steal LLM's \\ last layer}  & \makecell{WM-agnostic}  &  \ding{55} & \makecell{Extract LLM's \\ last  layer via API}\\
      \hline
      \hline
        \makecell{\textbf{$\delta$-\textsc{Steal} (Ours)} } & \makecell{WSteal LLM's \\  behaviors} & \makecell{WM-agnostic} &  \ding{51} &\makecell{Lightweight modifications \\(No modifications on LLMs)}  \\ 
        \bottomrule
    \end{tabular}
    \label{tab:Summary}
\end{table*}

\section{Related Work} \label{sec:relatedwork} 

There are two main types of attacks that threaten LLM intellectual property protection. First, \textit{watermark removal attacks} \citep{krishna2024paraphrasing,pan2024markllm,zhang2023watermarks} modify outputs to disrupt traceable watermark patterns using techniques like paraphrasing, synonym substitution, and sentence restructuring. While effective at weakening watermark detection, these methods often degrade text quality.
Second, \textit{stealing attacks} involve extracting watermark components, such as watermarking rules, vocabulary lists, or parts of a production LLM via APIs \citep{jovanovic2024watermark,carlini2024stealing,wu2024bypassing,birch2023model}. The goal is to replicate the behaviors of the service provider model and bypass watermark detectors, but existing methods are limited to specific watermark types and can only extract small portions of an LLM.

Table~\ref{tab:Summary} highlights key differences between $\delta$-\textsc{Steal} and related work. Our attack follows the second direction but focuses on stealing LLM behaviors  in a model-agnostic and watermark-agnostic manner. It only introduces a lightweight modification by adding LDP noise into the token embeddings, without changing the tokens, modifying LLM itself, or requiring model access. 
Furthermore, it can be easily applied to any LLM or watermark, enhancing its generalizability and scalability.

\section{Background}\label{sec:Background}

\subsection{Model Stealing Attacks}
Due to the high cost of training, LLMs are typically offered via APIs \citep{minthigpen}, but this does not prevent model stealing attacks \citep{chesterman2024good,zhang2023apmsa}. Such attacks can cause intellectual property violations, service replication, fee evasion, and system exploitation, underscoring the need for robust defenses. 
Model stealing attacks \citep{zhang2023apmsa,jovanovic2024watermark,carlini2024stealing} are also referred to as imitation attacks \citep{wallace2020imitation,xu2022student} or model extraction attacks \citep{krishna2019thieves,birch2023model}. 
To steal a model $\theta$, adversaries query the API to collect $N$ input-output pairs $\{x_i, y_i\}_{i=1}^N$ and train or fine-tune a surrogate model $\theta_{adv}$. The goal is to replicate the behavior of $\theta$ while bypassing watermark detectors.

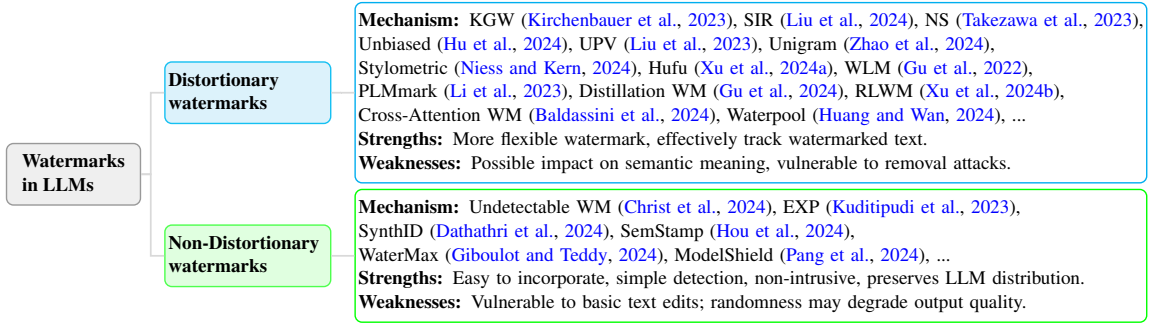
\begin{figure*}[t!]
  \centering
  \resizebox{\textwidth}{!}{%
    \begin{forest}
      forked edges,
      for tree={
        grow=east,
        reversed=true,
        anchor=base west,
        parent anchor=east,
        child anchor=west,
        base=center,
        rectangle,
        draw=hidden-draw,
        rounded corners,
        align=left,
        text centered,
        minimum width=4em,
        edge+={darkgray, line width=1pt},
        s sep=3pt,
        inner xsep=2pt,
        inner ysep=3pt,
        line width=0.8pt,
        ver/.style={rotate=90, child anchor=north, parent anchor=south, anchor=center},
      },
      where level=1{text width=10em,font=\normalsize,}{},
      where level=2{text width=10em,font=\normalsize,}{},
      where level=3{text width=11em,font=\normalsize,}{},
      where level=4{text width=7em,font=\normalsize,}{},
      [\textbf{Watermarks} \\ \textbf{in LLMs}, leaf-head
        [\textbf{Distortionary} \\ \textbf{watermarks}, blue-box-lv2, text width=8.3em
          [{\textbf{Mechanism:} KGW \citep{kirchenbauer2023watermark}, SIR \citep{liu2024semanticinvariantrobustwatermark}, NS \citep{takezawa2023necessary}, \\ Unbiased \citep{hu2024unbiased}, UPV \citep{liu2023unforgeable}, Unigram \citep{zhao2024provable}, \\ Stylometric \citep{niess2024stylometric}, Hufu \citep{xu2024hufu}, WLM \citep{guwatermarking}, \\ PLMmark \citep{li2023plmmark}, Distillation WM \citep{gu2023learnability}, RLWM \citep{xu2024learning}, \\ Cross-Attention WM \citep{baldassini2024cross}, Waterpool \citep{huang2024waterpool}, ... \\
          \textbf{Strengths:} More flexible watermark, effectively track watermarked text. \\
          \textbf{Weaknesses:} Possible impact on semantic meaning, vulnerable to removal attacks.}, 
          blue-box-lv4, text width=41em]
        ]
        [\textbf{Non-Distortionary} \\ \textbf{watermarks}, green-box-lv2, text width=8.3em
          [{\textbf{Mechanism:} Undetectable WM \citep{christ2024undetectable}, EXP \citep{kuditipudi2023robust}, \\ SynthID \citep{dathathri2024scalable}, SemStamp \citep{hou2023semstamp}, \\ WaterMax \citep{giboulot2024watermax}, ModelShield \citep{pang2024adaptive}, ... \\
          \textbf{Strengths:} Easy to incorporate, simple detection, non-intrusive, preserves LLM distribution. \\
          \textbf{Weaknesses:} Vulnerable to basic text edits; randomness may degrade output quality.}, 
          green-box-lv4, text width=41em]
        ]
      ]
    \end{forest}
  }
  \caption{Taxonomy of Watermarks in LLMs.}
  \label{figure:WMtaxonomy}
\end{figure*}

\subsection{Watermarking Techniques }
\label{sec:Watermark-based Defenses}
Recent work has shown that watermarks are effective in defending against model stealing attacks \citep{he2022cater,he2022protecting,kuditipudi2023robust}.
Given a prompt $x$, the service provider uses its model $\theta$  and a watermarking function $\mathcal{W}$ to generate a watermarked output $y^{wm} = \theta(x, \mathcal{W})$. A detector $\mathcal{D}(\cdot)$, based on statistical tests or classifiers \citep{kirchenbauer2023watermark,kuditipudi2023robust}, is then used to verify the presence of the watermark. By analyzing detection rates, service providers can set thresholds to identify whether a suspect model has stolen their model's behavior.

 Fig.~\ref{figure:WMtaxonomy} shows the  taxonomy of watermarks in LLMs, along  with   their advantages and disadvantages. Watermarks  can be classified into  distortionary and non-distortionary types. 
 \textit{Distortionary watermarks} modify LLM  weights or logits, affecting  its  functioning. Examples include logits-based watermarks \citep{kirchenbauer2023watermark,liu2024semanticinvariantrobustwatermark,zhao2024provable}, which change token generation logits, and training-based watermarks \citep{baldassini2024cross,gu2023learnability,xu2024learning}, which alter the  training process.  While effective for tracking, these watermarks may impact semantic meaning and are vulnerable to removal attacks \citep{carlini2024stealing,krishna2024paraphrasing,zhang2023watermarks}.
 \textit{Non-distortionary watermarks} inject watermarks without altering output distributions,   weights, or logits. Examples include sampling-based watermarks \citep{hou2023semstamp,kuditipudi2023robust}, which modify token or sentence sampling strategies, and prompt-based or multiple-output watermarks \citep{giboulot2024watermax,pang2024adaptive}, which generate multiple candidate outputs and select the one with the most identifiable watermark. These methods are harder to detect but can be vulnerable to text modifications and  may reduce reliability.

\subsection{Local Differential Privacy (LDP)}
LDP is a mathematically provable method for ensuring data privacy \citep{erlingsson2014rappor}, based on randomized survey responses 
\citep{warner1965randomized}, where noise is added to data to protect  confidentiality and prevent accurate inference of individual data.  

The LDP definition is as follows:

\begin{mydef}{} A randomized algorithm $\mathcal{A}$
satisfies $\epsilon$-LDP, if for any two inputs $x$ and $x'$, and for all possible outputs $\mathcal{O} \in Range(\mathcal{A})$, we have:
$Pr[\mathcal{A}(x) = \mathcal{O}] \leq e^{\epsilon} Pr[\mathcal{A}(x') = \mathcal{O}]$,
where $\epsilon$ is a privacy budget and $Range(\mathcal{A})$ denotes every possible output of  $\mathcal{A}$.  
\label{def: Different Privacy} 
\end{mydef}

The privacy budget $\epsilon$ controls the difference between the distributions induced by inputs $x$ and $x'$. A smaller $\epsilon$ results in a larger gap between the outputs and vice versa.

In this work, we apply LDP to perturb the token embeddings of adversaries' data,  bounding the differences between tokens and their associate  outputs  under LDP guarantees. We  employ the Laplace mechanism \citep{dwork2014algorithmic}, a commonly used LDP approach, to add Laplace noise. For all inputs $x$ and $x'$ in the domain of $\mathcal{A}$, where $d$ is the input dimension, the Laplace noise $Lap(0,\delta)$   has a zero mean and a noise scale $\delta$ calculated as follows:
\begin{equation}
    \small
    \delta= \frac{\max_{  \forall  x, x' \in R^{d}} \| \mathcal{A}(x) - \mathcal{A}(x') \|_1 }{\epsilon}
    \label{eq:LaplaceNoise}
\end{equation}

\section{$\delta$-\textsc{Steal}:  LLM  Stealing Attack with Local Differential Privacy}
\label{sec:certifiedLLM}

We introduce $\delta$-\textsc{Steal}, a novel model stealing attack that balances attack effectiveness and model utility. To achieve this, we address three fundamental questions: 
\textit{1) How can we optimize this trade-off?} \textit{2) Where should LDP noise be added for the best balance?} and \textit{3) How can we bound output differences to evade watermark detectors?} Our approach involves: \textit{1)} Leveraging  LDP concepts and  varying levels of LDP guarantees by adjusting the noise scale $\delta$ to better control the trade-off,  \textit{2)} Adding LDP noise to token embeddings  without modifying the tokens themselves, preserving model utility, and \textit{3)} Using LDP guarantees to bound the differences between the outputs generated by the adversaries and   by the service provider, bypassing watermark detectors to enhance attack success.

\subsection{Setting}
This work considers a service provider offering its LLM $\theta$ via an API, where users can query a prompt $x$ and receive the corresponding output $y$. To protect its intellectual property, the service provider injects a watermark, producing $y^{wm} = \theta(x, \mathcal{W})$ to users. Model stealing attacks then aim to replicate the behavior of the target LLM $\theta$ given a prompt $x$ in this black-box setting, where users have no access to the model's internal structure, parameters, or inference details (e.g., temperature, cached memory). This setting reflects practical deployments widely adopted by commercial providers.

\subsection{Threat Models}

\textbf{Adversary's goals.} The adversary aims to 1) replicate the service provider's model without direct access, 2) bypass watermark detectors, 3) obfuscate model ownership to hinder infringement claims, and 4) preserve utility comparable to the original. Achieving these goals enables the adversary to launch a competing service, causing potential revenue loss for the service provider's model $\theta$.

\textbf{Adversary's capabilities.} We assume an attacker who can query the service provider's LLM via API like any user and also has no knowledge of LLM's internal specifications. The LLM discloses no details in its outputs, ensuring no information leakage through the output $y$.

 \begin{figure*}[t]
      \centering    
      \includegraphics[scale=0.1]{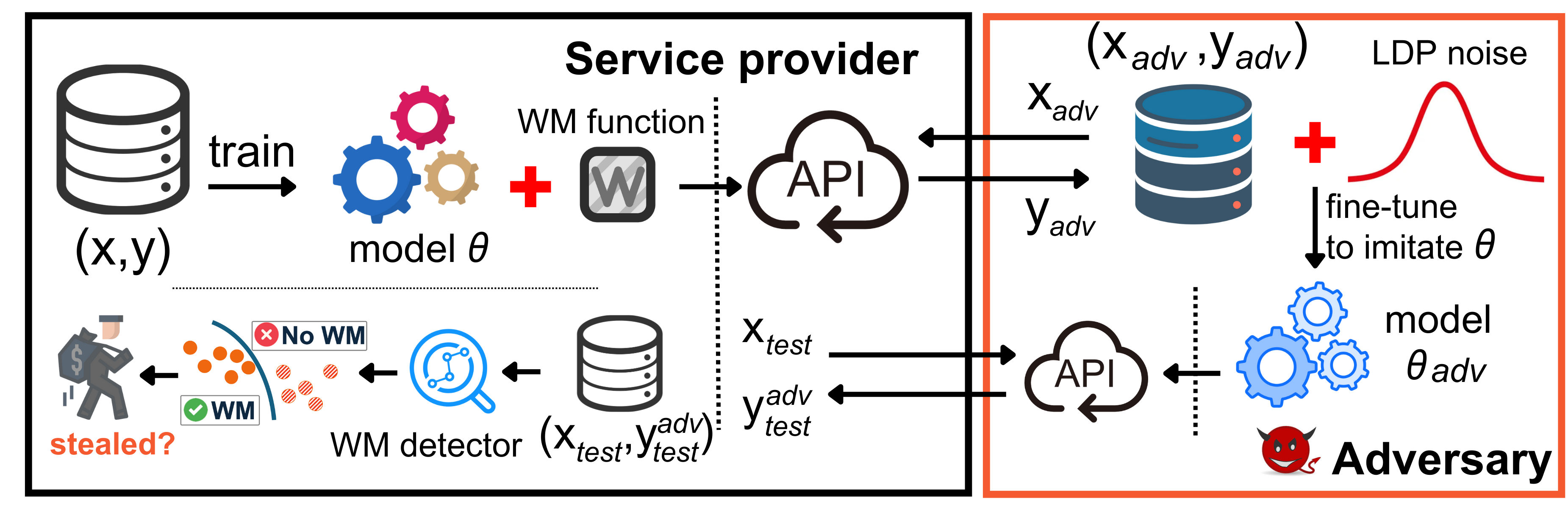} 
      \caption{System architecture of  $\delta$-\textsc{Steal}.} 
      \label{fig:overview}
 \end{figure*}

\subsection{System Operation of $\delta$-\textsc{Steal}}
Fig.~\ref{fig:overview} and Algorithm \ref{alg: LDPstealIP}  illustrate the system architecture and pseudo-code of our $\delta$-\textsc{Steal}.

\textbf{At the service provider}, the model $\theta$ is trained on proprietary data (Line 5) and deployed via an API or MLaaS platform.
To protect intellectual property, outputs are watermarked with $\mathcal{W}$ (Line 6). In addition, the service provider employs a watermark detector $\mathcal{D}$ to identify whether an arbitrary output $y_{adv}$ from a suspect model $\theta_{adv}$ contains a watermark; detection indicates that $\theta_{adv}$ was trained on watermarked data (Lines 8–11), enabling prompt detection of intellectual property violations.

\textbf{At the adversary}, the goal is to replicate the service provider's model $\theta$ while preserving high utility. To achieve this, the adversary queries the service provider's model $\theta$ with $N$ prompts $\{x_i\}_{i=1}^N$ and collects watermarked responses $y^{wm}_i = \mathcal{W}(\theta(x_i))$ (Line 15). These prompt-output pairs $\{x_i, y^{wm}_i\}_{i=1}^N$ are then used to fine-tune the adversary's model $\theta_{adv}$. By leveraging these correlations, the adversary can mimic the service provider model's functionality and behavior during training. 
Unlike conventional attacks \citep{zhang2023watermarks,pan2024markllm}, which typically modify tokens with synonyms but often compromise text coherence, $\delta$-\textsc{Steal} adds LDP noise to token embeddings without modifying the tokens themselves, thus effectively maintaining utility (Line 16). It is worth noting that the LDP noise is only added once before fine-tuning  $\theta_{adv}$. This prevents accumulation of the privacy budget $\epsilon$, ensuring tighter LDP guarantees on the output differences between $\theta$ and $\theta_{adv}$. 

\subsection{Bounding Output Differences with Local Differential Privacy  (LDP)}

\noindent{\textbf{Theoretical Bound.}}  
Watermark detectors exploit distinctive statistical patterns between the provider's model $\theta$ and the adversary's surrogate $\theta_{adv}$. An effective attack must obscure these patterns while preserving utility, a non-trivial trade-off. Appx.~\ref{Appx.A} shows that applying an $\epsilon$-LDP to embeddings bounds the change in output distributions by a factor of $\text{exp}^\epsilon$, thereby constraining watermark detectability. This provides a theoretical link between $\epsilon$, $\delta$ and the likelihood of successful evasion.  

\noindent{\textbf{Noise Calibration.}}  
Our choice of $\delta \in \{0.001,0.01,0.05,0.1\}$ is guided by the global sensitivity of embedding vectors detailed in Appx.~\ref{Appx.B}, capturing the transition from negligible perturbation to semantic distortion. Fig.~\ref{fig:LLM_sensitivity}b in Section~\ref{sec:varyingcomponents} further connects this to empirical sensitivity differences across LLMs, showing how embedding-level statistics (e.g., mean and variance) from different LLM profiles could inform $\delta$ selection to balance watermark evasion with utility preservation.  

\noindent{\textbf{Embedding-layer Noise}}. We inject noise at the embedding layer because it offers stronger protection and better utility retention than other deeper layers. First, output-level or latent-level noise is more vulnerable to query averaging and reconstruction attacks \citep{dwork2014algorithmic}, whereas adding noise to the embedding, we can hide the noise in model parameter though training, which would enhance the attack. Second, prior work shows that calibrating LDP on embeddings preserves semantics and yields favorable utility \citep{meisenbacher2024comparative,feyisetan2020privacy}. In addition, training-time noise at this layer also acts as a form of regularization \citep{bishop1995training}, allowing early distortion control while avoiding averaging at inference. Lastly, as highlighted, embedding statistics provide $\delta$ guideline, reinforcing why the embedding layer is the most effective place to inject noise.

\begin{algorithm}[t]
\small
\caption{$\delta$-\textsc{Steal} Algorithm}\label{alg: LDPstealIP}
\begin{algorithmic}[1]
\STATE \textbf{Inputs}: Service provider's  model $\theta$, adversary's model $\theta_{adv}$, $N$  adversary's prompts $\{x_i\}_{i=1}^N$,  $N_{test}$ service provider's testing prompts $\{x_j\}_{j=1}^{N_{test}}$, training iterations $T$, watermark $\mathcal{W}$, detector $\mathcal{D}$, loss $\mathcal{L}$
\STATE \textbf{Outputs}: Adversary's model $\theta_{adv}$ and watermark detector results $ \mathbb{I} ( \theta_{adv}, \theta)$ 
\STATE \textbf{At the Service Provider}:
\begin{ALC@g}
\STATE \textit{\textbf{Watermark Injection}}
\begin{ALC@g}
\STATE Initialize model parameters and fine-tune $\theta$ using training data  available at the service provider
\STATE Inject a watermark $\mathcal{W}$ to outputs  before releasing them to users:  $y^{wm} = \theta(x, \mathcal{W})$
\end{ALC@g}
\STATE \textit{\textbf{Watermark Detection}}
\begin{ALC@g}
\FOR{$j = 1, \ldots, N_{test}$}
\STATE Query the adversary's model to collect its outputs: $y_j = \theta_{adv}(x_j)$ 
 \STATE Check if $y_j$ is watermarked: $\mathbb{I}\big(\mathcal{D}(y_j)=1\big)$
\ENDFOR
\end{ALC@g}
\STATE Return $ \mathbb{I} ( \theta_{adv}, \theta)$ based on $\mathbb{I}\big(\mathcal{D}(y_j)=1\big)$ for all $\{y_j\}_{j=1}^{N_{test}}$
\end{ALC@g}
\STATE \textbf{At the Adversary}:
\begin{ALC@g}
\STATE Initialize model parameters $\theta_{adv}$ 
\FOR{$i = 1, \ldots, N$}
\STATE Query the service provider's model to collect its outputs: $y^{wm}_i =  \theta(x_i, \mathcal{W})$ 
 \STATE {\color{blue}Add Laplace LDP noise into token embeddings  with a noise scale $\delta$:\\ \quad $ \{\bar{x}_i, \bar{y}_i\}= \{x_i + Lap(0,\delta), y^{wm}_i + Lap(0,\delta) \}$ }
 \ENDFOR
\STATE Form a training set: $D_{adv} = \{\bar{x}_i, \bar{y}_i\}_{i=1}^N$ 
\FOR{$t = 1, \ldots, T$}
\STATE Randomly select a set of training samples $D_t \subseteq D_{adv} $   
\STATE $\theta_{adv}^{t} = \theta_{adv}^{t-1} - \eta \bigtriangledown_{\theta_{adv}} \mathcal{L} (\theta_{adv}^t, D_t)$  
\ENDFOR
\STATE Return $\theta_{adv} = \theta_{adv}^T$
\end{ALC@g}
\end{algorithmic} 
\end{algorithm} 

\section{Experiments}
\label{sec:experiments}
Our extensive experiments shed light on  \textbf{1)} $\delta$-\textsc{Steal}'s effectiveness against watermarks; \textbf{2)} Its impact on model utility, including text generation and downstream tasks; \textbf{3)} Comparisons with existing attacks; \textbf{4)} The trade-off between attack success and utility under LDP guarantees; and \textbf{5)} The influence of $\delta$-\textsc{Steal}'s components on this trade-off.

\subsection{Baselines}
We evaluate six state-of-the-art (SOTA) watermarks: \textbf{1)} \textit{KGW} \citep{kirchenbauer2023watermark}, which splits the vocabulary into green/red tokens and biases logits toward green ones; \textbf{2)} \textit{EXP} \citep{kuditipudi2023robust}, which maps sequence keys to tokens during sampling; \textbf{3)} \textit{SIR} \citep{liu2024semanticinvariantrobustwatermark}, which adjusts logits based on the semantics of previous tokens; \textbf{4)} \textit{SemStamp} \citep{hou2023semstamp}, which accepts sentences mapped to valid semantic regions; \textbf{5)} \textit{TW} \cite{yang2023watermarking}: A binary text watermark that replaces tokens with context-aware synonyms, guided by a Bernoulli-based random encoding; and \textbf{6)} \textit{DeepTextMark} \cite{munyer2024deeptextmark} (referred to as \textit{DTM}), which substitutes tokens with synonyms using Word2Vec \cite{mikolov2013efficient}. 
Each method includes its own watermark detector and watermarked output under attack-free environment is denoted as the \textit{Baseline}.

In addition, we compare $\delta$-\textsc{Steal} with three SOTA watermarking attacks: \textbf{1)} \textit{WMremoval} \citep{zhang2023watermarks}, which paraphrases outputs while using a quality oracle to preserve fluency; \textbf{2)} \textit{Dipper} \citep{krishna2024paraphrasing},  which paraphrases the outputs through context reordering and lexical changes; and \textbf{3)} \textit{Substitution}, adapted from \citep{pan2024markllm}, which replaces words with WordNet synonyms \citep{miller1995wordnet}. An LLM output without watermarks or attacks is denoted as the \textit{Original}.

\subsection{Dataset and Model Configurations}
We evaluate $\delta$-\textsc{Steal} on text generation and downstream tasks. \textit{For text generation}, we randomly select $10,000$ training samples and $2,000$ test samples from the C4 dataset \citep{dodge2021documentinglargewebtextcorpora}, using the first $200$ tokens as prompts and generating up to $200$ tokens. This setup applies to Original and Baseline. \textit{For the downstream task}, we use the massive multi-task language understanding MMLU \citep{hendrycks2021measuringmassivemultitasklanguage}, a multiple-choice benchmark across 57 subjects of varying difficulty.

We conduct  experiments on two LLMs, including LLaMA-2 7B \citep{touvron2023llama2openfoundation} and Mistral 7B \citep{jiang2023mistral7b}, as the Original models. We  employ a Laplace mechanism \citep{dwork2014algorithmic}, widely used in LDP, to add noise to the token embeddings. We vary the Laplace noise $Lap(0,\delta)$  with different   noise scales  $\delta \in \{0.001, 0.01, 0.05, 0.1\}$. These noise scales are associated with $\epsilon$ values of $\{300, 30, 6, 3\}$ for the LLaMA-2 and $\{50, 5, 1, 0.5\}$ for the Mistral.  Details on how $\delta$ is computed from noise scales are provided in Appx.~\ref{Appx.B}. We fine-tune adversary models using LoRA \citep{hu2022lora} with the Adam optimizer on 10 epochs with a learning rate of $10^{-5}$.

\subsection{Evaluation Metrics}
A model stealing attack is considered successful if it achieves \textit{high attack success rates} while \textit{preserving utility}, ensuring strong performance on both the main task (e.g., text generation) and downstream tasks (e.g., MMLU). To evaluate $\delta$-\textsc{Steal}, we examine three key aspects. \textit{First}, for effectiveness, we compute the attack success rate (AttackSR) as follows:
\begin{equation}
\small
    \text{AttackSR} = 1- \frac{ \sum_{i=1}^{N_{test} }\mathbb{I} \Big(\mathcal{D}(\theta_{adv}(x_i))=1 \Big)}{N_{test}}
    \label{eq:IPcheckerEq}
\end{equation}

\noindent where $N_{test} $ is the number of test samples located at the service provider and $\mathcal{D}(\theta_{adv}(x_i))$ represents the watermark detector, such that  $\mathcal{D}(\theta_{adv}(x_i))=1$ if the output of $\theta_{adv}$ given  $x_i$, denoted as $\theta_{adv}(x_i)$, is  watermarked and $\mathcal{D}(\theta_{adv}(x_i))=0$ otherwise. In addition,  $\mathbb{I}$ is the indicator function, where  $\mathbb{I}(x)=1$ if $x$ is True,  and  $\mathbb{I}(x)=0$ otherwise. Eq.~\ref{eq:IPcheckerEq} defines AttackSR, which measures the failure rate of watermark detection on the adversary's outputs; higher values indicate better evasion. \textit{Second}, we assess utility using: 1) Perplexity (PPL) for text generation, and 2) Average accuracy on MMLU \citep{hendrycks2021measuringmassivemultitasklanguage}. \textit{Third}, we further perform qualitative analysis through side-by-side visualization of prompt-output pairs across watermarks and attacks on different LLMs.

\begin{figure}[t]
 \centering
\subfigure[AttackSR (LLaMA-2)]{\includegraphics[scale=0.32]{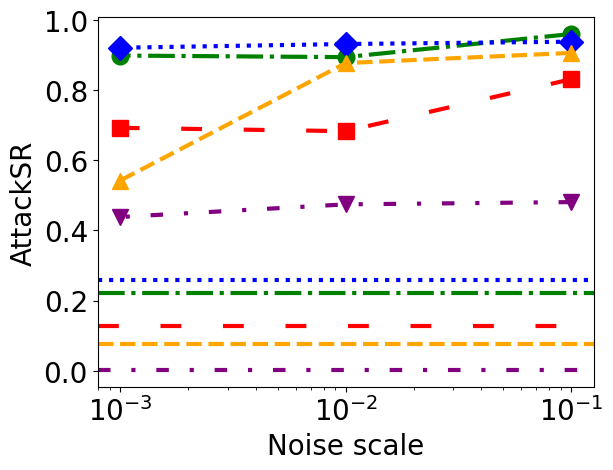}}
    \subfigure[AttackSR (Mistral)]{\includegraphics[scale=0.32]{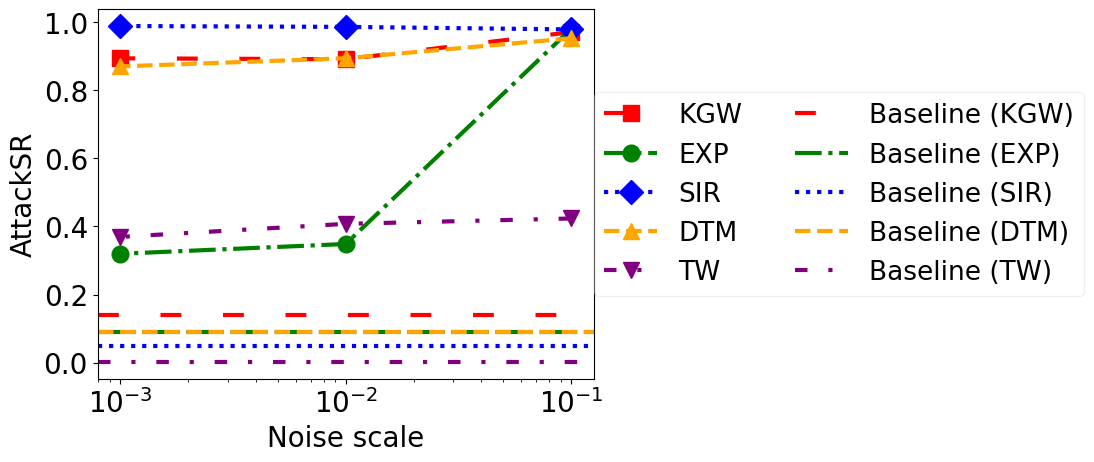}}\\
    \subfigure[Perplexity (LLaMA-2)]{\includegraphics[scale=0.32]{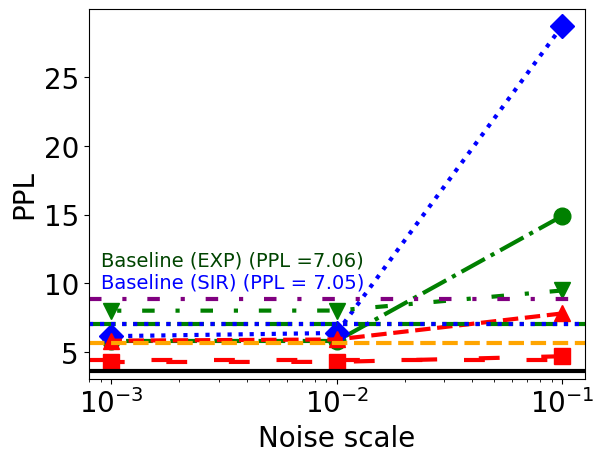}}
    \subfigure[Perplexity (Mistral)]{\includegraphics[scale=0.32]{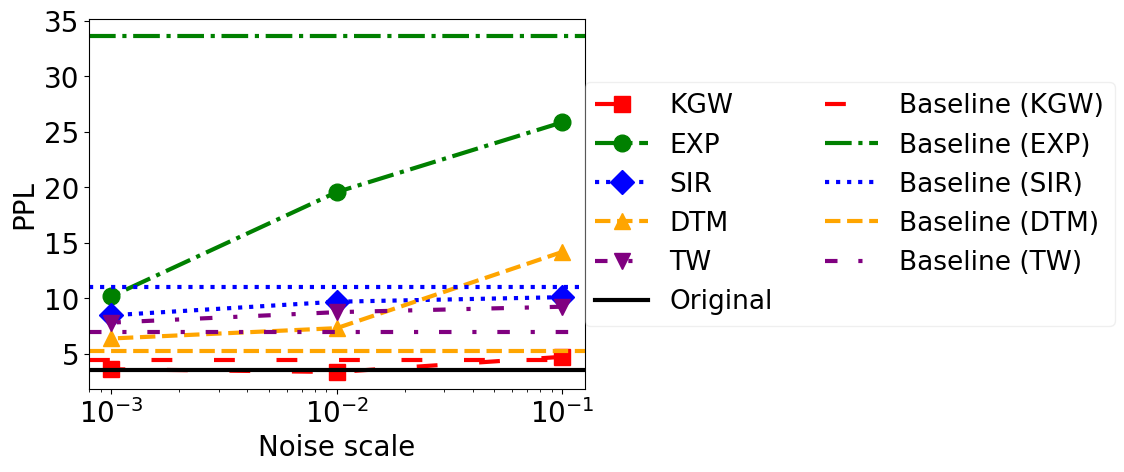}}
\caption{$\delta$-\textsc{Steal} performance on different watermarks.} 
\label{fig:combined_watermark_evasion}
\end{figure}

\subsection{Experimental Results}  \label{sec:experiment_result}

\subsubsection{ $\delta$-\textsc{Steal} against Existing watermarks} 

In Fig.~\ref{fig:combined_watermark_evasion},  $\delta$-\textsc{Steal} demonstrates effectiveness across watermarks, LLMs, and noise scales, achieving high AttackSR and low PPL (where lower is better). For example, with LLaMA-2 at $\delta = 0.001$, $\delta$-\textsc{Steal} yields $69.28$\% AttackSR and $4.29$ PPL on KGW, but achieves $89.90$\% and $92.13$\% AttackSR with $5.82$ and $6.17$ PPL on EXP and SIR, respectively. These results show that $\delta$-\textsc{Steal} is more evasive with SIR, achieving a higher AttackSR with minimal impact on  PPL.  In addition, as the noise scale $\delta$ increases, AttackSR improves while PPL slightly worsens yet remains close to the Baseline and Original outputs. For instance, with   KGW on  Mistral, as  the noise scale $\delta$ increases from $0.001$  to $0.1$,  AttackSR rises from  $89.28\%$ to  $96.95\%$  while PPL slightly increases from $3.61$ to $4.73$. We observe similar trends with EXP and SIR,  showing  increases of $1.72\%-6.14\%$  in AttackSR and $9.08-23.41$ in PPL values. Additional results from DTM and TW remained consistent, achieving high AttackSR and minimal utility degradation, further confirming $\delta$-\textsc{Steal}'s effectiveness.

Furthermore, the performance does not exhibit a linear trade-off between AttackSR and PPL. This non-linear response is expected due to the inherent randomness of noise injection and the discrete nature of text. If AttackSR is already near its maximum, further increases in noise will not significantly improve evasion, resulting in flat regions. Similarly, for certain watermarks, moderate noise may be sufficient to destroy the detector, so AttackSR rises quickly while PPL remains relatively stable over a range of $\delta$. These patterns reflect threshold effects rather than linear trade-offs.

To validate our work, we conducted statistical significance testing on a representative setting of KGW under LLaMA-2 with $\delta=0.01$. The experiment was repeated five times, comparing the mean AttackSR with the reported $68.34\%$ (Table~\ref{tab:kgw-significance}). The observed mean of $68.94\%$ is close to the reported value, and a one-sample $t$-test yields a $p$-value of $0.486$, indicating no significant difference. These results confirm the consistency and reliability of our findings beyond a single trial.

\begin{table}[h]
\small
\centering
\caption{AttackSR significance testing of KGW on LLaMA-2 with $\delta=0.01$.}
\label{tab:kgw-significance}
\begin{tabular}{lcccccc}
\toprule
Reported AttackSR & Trial 1 & Trial 2 & Trial 3 & Trial 4 & Trial 5 & $p$-value \\
\midrule
68.34\% & 68.85\% & 69.00\% & 68.10\% & 67.01\% & 71.75\% & 0.486 \\
\bottomrule
\end{tabular}
\end{table}

Intuitively, $\delta$-\textsc{Steal} introduces noise into token embeddings, and as the noise scale increases (i.e., higher values of $\delta$), the modifications become more substantial. This makes it increasingly difficult for watermark detectors to detect the watermarks, thereby improving   AttackSR. \textit{In distortionary watermarks} such as  KGW or SIR the generated watermarked tokens depend on preceding context. Perturbing token embeddings can alter token selection, causing noise accumulation across tokens and disrupting the watermark signature. Similarly, for \textit{non-distortionary watermarks} such as EXP, watermarked tokens are chosen via a predefined sampling process while preserving the probability distribution. However, the introduction of LDP noise into token embeddings causes deviations in the rule-based token sampling process. This results in inconsistencies in token choices and  weakens the injected watermark patterns, reducing detectability of watermark detectors.

Despite these perturbations, $\delta$-\textsc{Steal} highly preserves model utility for two  reasons. First, the perturbed outputs stay close to the original distribution while the training pipeline remains unchanged, allowing the surrogate $\theta_{adv}$ to capture semantic and syntactic patterns of the provider's model. Second, since the noise is added only during fine-tuning in a controllable LDP-preserving manner, the core learning signal remains intact. This allows the stolen model's outputs to retain fluency, coherence, and task performance, even while obfuscating  watermark patterns. As a result, $\delta$-\textsc{Steal} empirically achieves a strong balance between watermark evasion and model functionality, making it highly effective for stealing proprietary LLM behavior without degrading output quality.

\begin{figure}[H]
    \centering
    \begin{minipage}{\linewidth}
        \subfigure[LLaMA-2 with KGW]
        {\includegraphics[scale=0.3]{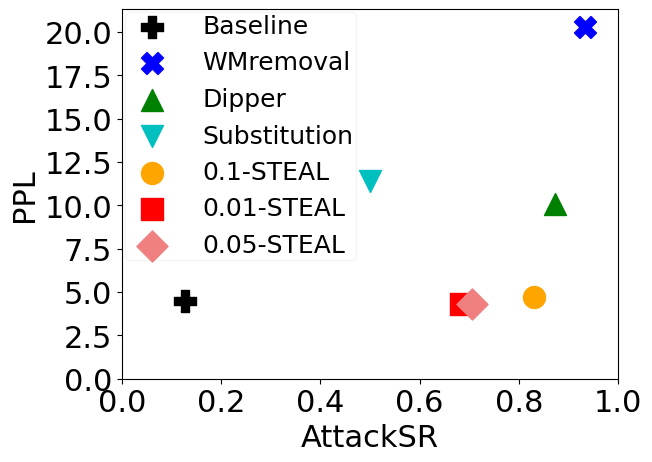}}\hfill
        \subfigure[LLaMA-2 with EXP] 
        {\includegraphics[scale=0.3]{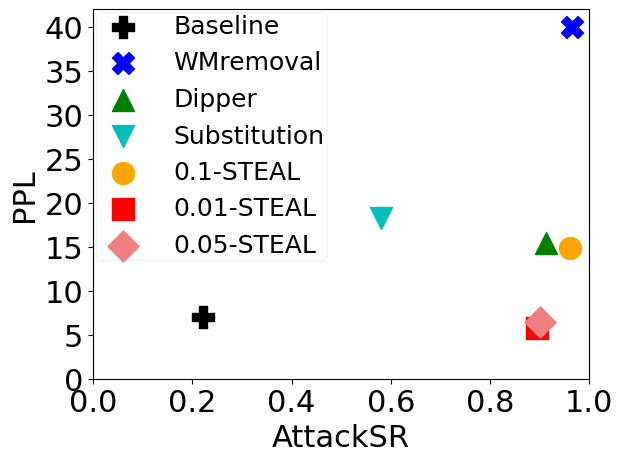}}\hfill
        \subfigure[LLaMA-2 with SIR] 
        {\includegraphics[scale=0.3]{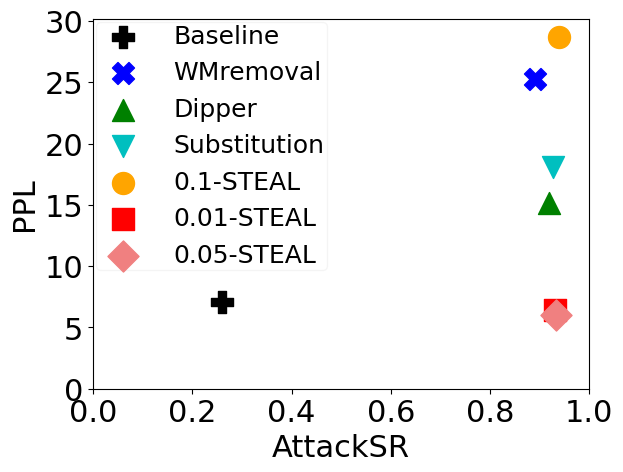}}
    \end{minipage}
    \hspace{0.001cm} 
    \begin{minipage}{\linewidth}
        \subfigure[Mistral with KGW] 
        {\includegraphics[scale=0.3]{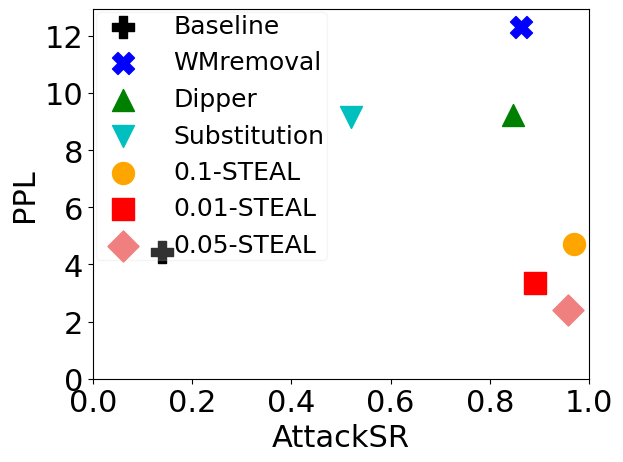}}\hfill
        \subfigure[Mistral with EXP] 
        {\includegraphics[scale=0.3]{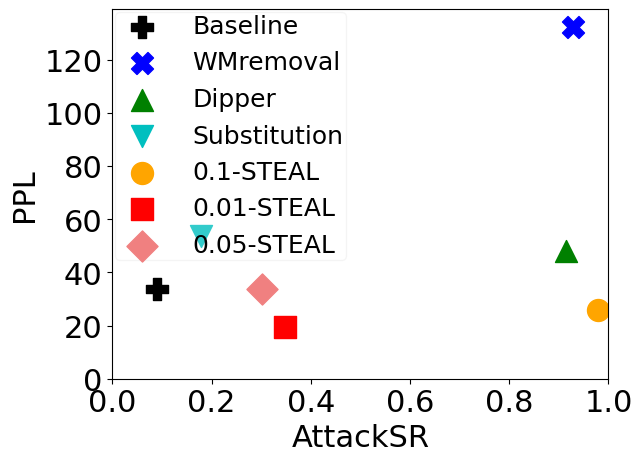}}\hfill
        \subfigure[Mistral with SIR]
        {\includegraphics[scale=0.3]{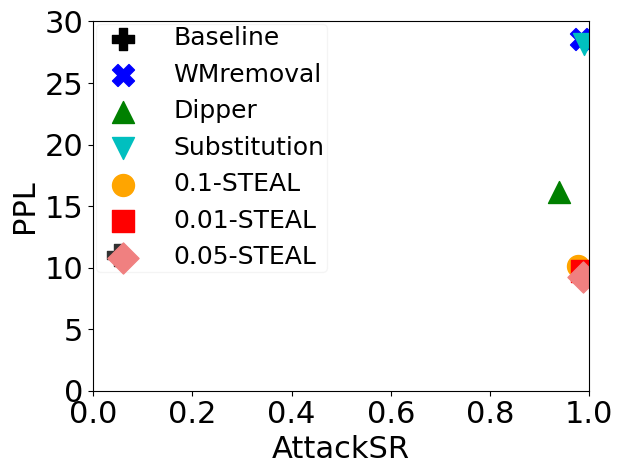}}
    \end{minipage}
    \caption{AttackSR and Perplexity results.   (Best attacks are in the bottom-right.)}
    \label{fig:compare_with_others}
\end{figure}
\vspace{-1.5em}

\subsubsection{ Comparison of $\delta$-\textsc{Steal} and Existing Attacks} 
Fig.~\ref{fig:compare_with_others} compares the performance of $\delta$-\textsc{Steal} with other attacks (WMremoval, Dipper, and Substitution). Notably, a higher AttackSR and lower PPL indicate a better attack; therefore, attacks appearing in the bottom-right corner are considered more effective.
As shown, $\delta$-\textsc{Steal}  achieves a high AttackSR while maintaining low PPL, comparable to that of the Baselines. For instance, with $\delta = 0.1$, $\delta$-\textsc{Steal} achieves an AttackSR of $83.14\%$ and a PPL of $4.71$, compared to $4.46$ of LLaMA-2 with KGW Baseline. Similarly, for Mistral with KGW,  $\delta$-\textsc{Steal}   achieves an AttackSR of $96.95\%$   at  a PPL of $4.73$, compared to the Baseline PPL of $4.43$. Meanwhile, other attacks usually exhibit lower values of AttackSR but much higher values of PPL, indicating a substantial impact on the watermarked outputs, reduced attacked output quality. For instance, WMremoval achieves a high AttackSR of $93.44\%$ for LLaMA-2 with KGW but raises PPL to $20.31$ from the Baseline of $4.46$. It is also computationally expensive, requiring paraphrasing and quality checks for every token to generate non-watermarked outputs. Dipper encounters similar issues, reaching an AttackSR of $87.37\%$ at the cost of $10.09$ PPL while significantly degrading model utility on downstream tasks (Table \ref{tab:model_utility_comparison}). Substitution attack performs worst, showing high PPL or low AttackSR across settings due to random token replacements that disrupt semantics and reduce utility.

\subsubsection{$\delta$-\textsc{Steal} on  Downstream Tasks}

\noindent
\begin{minipage}{0.42\textwidth}
\centering
\captionof{table}{MMLU accuracy across LLMs and watermarks.}  
\label{tab:model_utility_comparison}
\small
\begin{tabular}{|c|c|c|c|c|c|}
    \hline
    \multicolumn{2}{|c|}{\shortstack{Accuracy  (\%)} } & \shortstack{LLaMA-2} & \shortstack{Mistral} \\ 
    \hline
    \multicolumn{2}{|c|}{Original} & 46.70 & 58.90 \\ 
    \hline
    \multirow{3}{*}{KGW} & Baseline & 44.50 & 58.60 \\ 
                         & $0.001$-\textsc{Steal} & 44.10 & 42.30 \\ 
                         & $0.01$-\textsc{Steal} & 43.00 & 39.30 \\
    \hline
    \multirow{3}{*}{EXP} & Baseline & 45.90 & 58.70 \\ 
                         & $0.001$-\textsc{Steal} & 45.50 & 47.70 \\ 
                         & $0.01$-\textsc{Steal} & 43.30 & 29.20 \\
    \hline
    \multicolumn{2}{|c|}{Dipper} & \multicolumn{2}{c|}{9.60} \\ 
    \hline
\end{tabular}
\end{minipage}\hfill%
\begin{minipage}{0.528\textwidth}
Table~\ref{tab:model_utility_comparison} compares MMLU performance under three settings: \textit{1)} the \textit{Original}, \textit{2)} \textit{Baseline}, and \textit{3)} $\delta$-\textsc{Steal} attack applied to the Baseline with noise scales $\delta \in [0.001, 0.01]$. Overall, $\delta$-\textsc{Steal} maintains model utility with only a subtle drop in accuracy compared with the Original and Baseline.  For LLaMA-2, $\delta$-\textsc{Steal} maintains strong performance across  all watermarks, with a small accuracy drop of $1.5-2.6\%$ at   $\delta  = 0.01$. Meanwhile,   Mistral exhibits greater sensitivity, showing a more significant accuracy drop up to $29.5\%$  at   $\delta  = 0.01$ and $16.30\%$ at   $\delta  = 0.001$. 
\end{minipage}
\vspace{1.0em}

The MMLU task is characterized by low entropy and more deterministic text, since outputs are multiple-choice answers, thus making it highly sensitive to noise. Even small perturbations can lead to noticeable drops in performance, especially in models like Mistral. In contrast, Dipper, the most effective attack in the early experiments, performs poorly on this task, achieving only $9.6\%$ accuracy. This is due to its aggressive paraphrasing process, which significantly alters the semantics of the text and disrupts the model's ability to select the correct multiple-choice answers. These observations further emphasize $\delta$-\textsc{Steal}'s effectiveness in maintaining model utility unlike other attacks.

\subsubsection{Trade-off between Attack Effectiveness and Model Utility in $\delta$-\textsc{Steal}}

Throughout all experiments, we observe a trade-off between  attack effectiveness and model utility. As the noise scale $\delta$ increases, AttackSR improves, while model utility decreases. 
Intuitively, increasing the noise scale $\delta$ introduces greater perturbations to the token embeddings.  This increases the difference between outputs generated by the service provider and those generated by the adversary, making it more challenging for watermark detectors to detect the presence of watermarks, thereby improving the AttackSR. However, the added noise can distort text quality, making it crucial to choose an optimal noise scale $\delta$ that could efficiently balance attack effectiveness and model utility. In this work, we present empirical results with $\delta \in [0.001, 0.1]$, and discuss its effect on the embedding space in section \ref{sec:varyingcomponents}, leaving the theoretical analysis of optimal noise for future work.

\begin{table*}[t!]
\centering
\scriptsize 
\caption{Prompts and watermarked outputs w/wo $\delta$-\textsc{Steal}. \hl{Green} is \hl{similarities}, \hlred{red} is   \hlred{differences}.}
\begin{tabular}{|p{0.175\linewidth}|p{0.285\linewidth}|p{0.30\linewidth}|p{0.05\linewidth}|p{0.045\linewidth}|}
\hline
\textbf{Prompt} &  \textbf{Watermarked output (W)} & \makecell{\textbf{$\delta$-Steal} \textbf{($\delta = 0.01$) output (A) } } & \textbf{PPL(W)} & \textbf{PPL(A)} \\
\hline
\raggedright ... do that for the MCD polls, '' said a senior party leader. Another reason behind not targeting Modi in the MCD polls is the massive victory the B... & 
\raggedright \textbf{{\color{blue} KGW:}}  \hl{JP} \hlred{achieved} \hl{in the 2014 Lok Sabha} \hlred{polls} in \hlred{Delhi} , where the party won all seven seats. `` \hl{We have to} \hlred{take into account the ground reality} and the political equation in the city.  \hlred{We can not ignore} the \hl{BJP 's victory in the 2014 Lok Sabha} [...continues] & 
 
\raggedright \textbf{{\color{blue} KGW:}}  \hl{JP} \hlred{registered} \hl{in the 2014 Lok Sabha} \hlred{elections} in \hlred{the city}. `` \hl{We have to} \hlred{convince the people of Delhi} that why we are the best option to govern the city. \hlred{We have to show} that the \hl{BJP 's victory in the 2014 Lok Sabha} elections was a [...continues]  & 
3.33 & 3.67 \\
\hline
\raggedright … Daniels, who was born Stephanie Clifford, was paid 130,000 by Cohen after she signed a nondisclosure agreement barring her from talking about her alleg... & 
\raggedright  \textbf{{\color{blue} EXP:}}   \hl{ed affair with Trump}. Trump \hlred{denies the allegation}. \hl{Cohen} has admitted making \hl{the payment to Daniels}, which he said was done to protect \hlred{Trump' s campaign from the allegations}. It' s possible that the reimbursement \hl{payment} \hlred{revealed} in the \hl{financial disclosure} may [...continues] & 
\raggedright  \textbf{{\color{blue} EXP:}}  \hl{ed affair with Trump}. \hl{Cohen} recently revealed that \hlred{Trump personally reimbursed him} for \hl{the payment to Daniels}. As a result of the payment to Cohen, the government ethics office \hlred{sent a reminder to Trump} that \hlred{he must disclose} in his annual \hl{financial disclosure} report any \hl{payments} made [...continues] &
4.12 & 5.05 \\
\hline
\raggedright … other shows, it's a GoPro on a windshield, '' Foley said referring to Ride Along. `` I think if this was a show that was on... & 
\raggedright  \textbf{{\color{blue} SIR:}}  \hlred{a traditional} \hl{television} platform, \hl{we would be able to do more with it}. As it stands, it's a while  \hlred{we do any new episodes of Holy Folesy!} '' \hl{Foley also} \hlred{addressed} \hl{why his daughter Noelle is n't pursuing} a career  [...continues] & 
\raggedright \textbf{{\color{blue} SIR:}}  \hlred{a major network} \hl{television}, \hl{we would be able to do more with it}. I think we would have a bigger budget and \hlred{be able to do some cool things with it}. '' \hl{Foley also} \hlred{talked about} \hl{why his daughter Noelle is n't pursuing} an [...continues] &
8.12 & 8.23 \\
\hline
\end{tabular}%
\label{table:output_comparison}
\end{table*}
\setlength{\textfloatsep}{10pt}

\subsubsection{Semantic Preservation of $\delta$-\textsc{Steal}  Outputs }

Table \ref{table:output_comparison} presents different watermarked baseline examples of LLaMA-2. Additional examples are  provided in Table \ref{tab:3a} (Appx.\ref{Appx.C}). 
We observe that  $\delta$-\textsc{Steal} effectively preserves semantic meaning compared to the Baselines, as indicated by the large green portions in text, while the red portions, despite textual changes, retain the original content. For instance, the $\delta$-\textsc{Steal} attack on KGW introduces slight modifications while preserving the overall semantic meaning, replacing ``elections in the city'' with ``polls in Delhi'' (based on earlier context mentioning Delhi) or ``We have to show that the BJP's victory'' with ``We cannot ignore the BJP's victory.''  Examples with EXP and SIR follow a similar pattern. For instance, in the  EXP example, given a financial disclosure context, ``Trump denies the allegation'' is modified to ``Cohen recently revealed that Trump personally reimbursed him,'' subtly  
shifting the focus but retaining the core idea.  In the SIR example,    ``traditional television platform'' is changed to ``major network television'' without significantly changing the meaning.  

\subsubsection{Varying Components of $\delta$-\textsc{Steal}}\label{sec:varyingcomponents}

\textbf{Varying LDP Levels.} Table \ref{tab:model_utility_comparison}  and Fig.~\ref{fig:combined_watermark_evasion} show that as the noise scale $\delta$ increases,  AttackSR  improves while PPL values  rise.  For instance, in text generation, with $\delta=0.001$, $\delta$-\textsc{Steal} achieves $69.28\%$ AttackSR and $4.29$ PPL under LLaMA-2 with KGW. When   $\delta$ increases to $0.1$,   AttackSR and PPL also increase to $83.14\%$ and $4.71$ respectively. Similarly, in the MMLU downstream task, at $\delta=0.001$, $\delta$-\textsc{Steal} achieves $44.10\%$ in average accuracy, but the performance drops to $43.00\%$ as $\delta$ is increased to $0.01$. This observation is consistent with different watermarks and LLMs.  

\begin{figure}[t]
 \centering
 \subfigure[MMLU Accuracy  (Mistral)]{\includegraphics[scale=0.34]{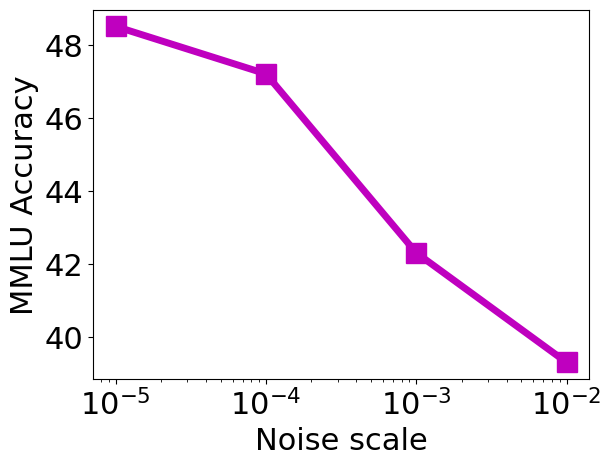}}\hspace{0.1cm} 
\subfigure[Embedding Distribution across LLMs]{\includegraphics[scale=0.29]{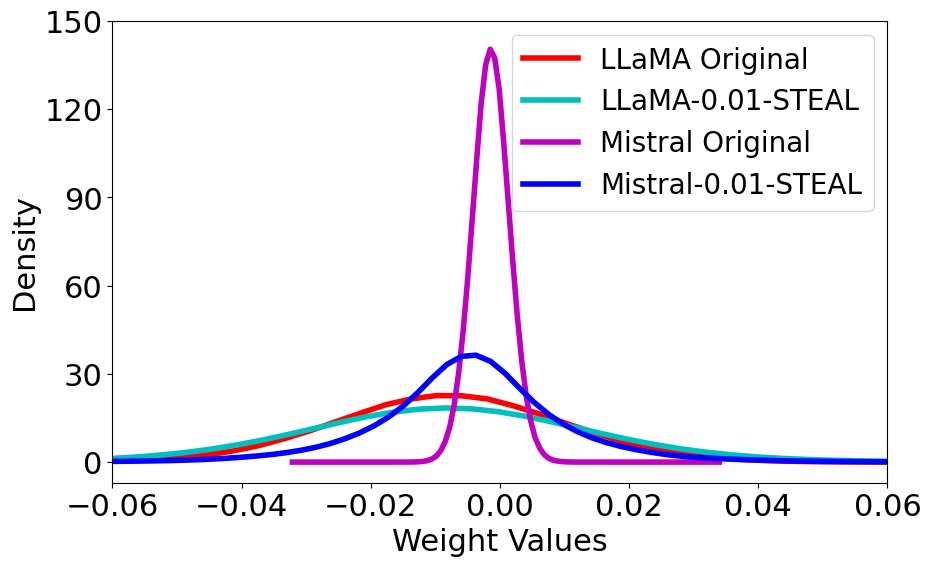}}
\hfill
\caption{MMLU accuracy and Sensitivity of different LLMs.} 
\label{fig:LLM_sensitivity}
\end{figure}

\textbf{Varying LLMs.} Table \ref{tab:model_utility_comparison} and Figs.~\ref{fig:combined_watermark_evasion}, \ref{fig:compare_with_others} exhibit similar trade-offs between attack effectiveness and   utility across watermarks and LDP levels across LLMs. However, Mistral demonstrates larger performance gaps across noise scales, whereas those of LLaMA-2 remains relatively stable. For instance, in Table \ref{tab:model_utility_comparison}, the average accuracy of MMLU using LLaMA-2 only drops by $0.8\%$ between noise scales of $0.01$ and $0.001$, while Mistral's average accuracy drops by $3.0\%$. Fig.~\ref{fig:LLM_sensitivity}a further shows that Mistral maintains high MMLU accuracy at smaller noise scales, achieving results comparable to the Baseline. At $\delta = 10^{-4}$, $\delta$-\textsc{Steal} on Mistral achieves $85.99$\% AttackSR with a low PPL of $4.44$.


 The performance gap between LLaMA-2 and Mistral under $\delta$-\textsc{Steal} stems from differences in their embedding layer weight distributions. In Fig.~\ref{fig:LLM_sensitivity}b, Mistral's embedding weights follow a sharply peaked distribution with a significantly smaller standard deviation of $0.0027$, while LLaMA-2's distribution is flatter, with a standard deviation of $0.01681$, approximately $6.22 \times$ higher. As a result, Mistral is more sensitive to perturbations, and applying the same noise scale to this LLM would cause greater distortion.  
 Therefore, analyzing the embedding weight distribution is a practical approach to guide the choice of noise scale $\delta$ and to anticipate its impact on attack performances.

SemStamp results are also reported in Appx.~\ref{Appx.C}. We used the authors’ pre-trained sentence embedder without fine-tuning, which is not well-suited for our setup. While fine-tuning a stronger embedder would likely improve detection, it is beyond the scope of this work. Nonetheless, the results provide useful insights into SemStamp's performance under these conditions.

\begin{table}[t]
\centering
\small
\caption{Effect of adversary training data size on AttackSR and PPL (LLaMA-2, KGW, $\delta=0.01$).}
\label{tab:scaling_attack}
\begin{tabular}{lccccccc}
\toprule
\textbf{Training size} & 100 & 1,000 & 5,000 & 10,000 & 20,000 & 50,000 & 100,000 \\
\midrule
\textbf{AttackSR (\%)} & 97.95 & 89.94 & 78.30 & 68.33 & 62.58 & 56.08 & 45.30 \\
\textbf{PPL}           & 4.54  & 4.36  & 4.56  & 4.30  & 5.09  & 5.14  & 5.17  \\
\bottomrule
\end{tabular}
\end{table}

\subsubsection{Varying Training Data Size}
Our experiments assume an attacker with a maximum query budget of $10^{4}$, which is  substantial and potentially costly in practice. However, an adversary could accumulate more prompts over time to train a local model. To study this, we vary the number of training samples   from $100$ to $100,000$ using LLaMA-2 with KGW and a noise scale of $0.01$. In Table \ref{tab:scaling_attack}, as the training size increases,  the AttackSR decreases. This is because  larger training datasets may lead the adversary model to overfit watermark-related patterns, making it more detectable.
These results highlight an important caution, in which more data does not necessarily equate to more effective attacks and underscore the need for overfitting-resistant training strategies in large-scale deployments of $\delta$-\textsc{Steal}.

\section{Discussion} \label{sec:discussion}
\noindent{\textbf{Novelty and Contribution.}}  While our work has limited novelty in terms of algorithmic mechanics, its contribution lies in being the first to integrate LDP theory with model stealing as a lightweight and empirically effective watermark evasion method. Prior studies on LDP have not considered this threat model, and we view our results as opening a new direction for the watermark security research.

\noindent{\textbf{Theoretical Limitations.}} Our framework does not fully formalize the link between $\epsilon$-LDP guarantees and the disruption of watermark-specific statistical features. The present work relies mainly on empirical evidence and intuition about embedding-layer noise to break detector correlations. Although Appx.~\ref{Appx.A} provides a bound on distributional shifts under $\epsilon$-LDP, it is not a formal proof that watermark detectability is eliminated. Establishing rigorous theoretical links between $\epsilon$, $\delta$ parameters and bounded watermark detection remains an important direction for future research.

\noindent{\textbf{Practical Deployment and Scalability.}} We acknowledge that real-world deployment may introduce additional complexity, such as diverse model architectures, varying query patterns, and adaptive watermarking defenses. However, the initial results from $\delta$-\textsc{Steal}, showing high attack success rates with minimal utility loss, suggest practical feasibility in deployment-like settings. While finer-grained $\delta$ values could provide more detailed insights, they would significantly increase training cost; the noise levels were carefully calibrated using embedding sensitivity, capturing the range from negligible perturbation to semantic distortion. In addition, leveraging efficient LoRA fine-tuning with modest hardware, $\delta$-\textsc{Steal} remains scalable to larger models and datasets.

\noindent{\textbf{Ablation on Noise Location.}} To better understand the effect of noise injection location, we conduct a study comparing three strategies on LLaMA-2 with KGW at $\delta=0.01$ on (1) noisy embeddings during fine-tuning ($\delta$-\textsc{Steal}), (2) noisy pre-logits, and (3) noisy embeddings applied at inference. Adding noise only at inference produces the lowest utility (highest PPL $5.27$) and the lowest AttackSR ($67.69$\%), since it directly affects model performance at deployment. Injecting noise at the pre-logits layer raises AttackSR by nearly $9$\% ($74.75$\%) but also increases PPL by $13$\% (PPL $4.86$), reflecting a less favorable utility–robustness balance. By contrast, $\delta$-\textsc{Steal} setting achieves the best trade-off, confirming the effectiveness of embedding-layer perturbation. Details are  reported in Appx.~\ref{Appx.D}.

\section{Conclusion}\label{sec:Conclusion}

This study introduces $\delta$-\textsc{Steal}, a novel model stealing attack that leverages LDP to bypass watermark detectors. By applying LDP to token embeddings without altering the tokens themselves, $\delta$-\textsc{Steal} effectively preserves model utility. In addition, we bound the differences  between watermarked and non-watermarked outputs, making it difficult for the service provider to distinguish whether the adversary's model was trained on the service provider’s watermarked data, thereby evading detection. We show that even with small noise, $\delta$-\textsc{Steal} can evade watermark detectors with a high AttackSR, reaching up to 96.95\% and an average of over 80\%, while maintaining model utility similar to the original model without attacks. Furthermore,  $\delta$-\textsc{Steal}  outperforms existing attacks across different watermarks,  LDP levels,  and LLMs,  enhancing its practical applicability in various scenarios.

Although our empirical analysis quantitatively shows the connection between the  injected noise scale $\delta$ and attack effectiveness, establishing formal theoretical connections remains a challenging and open problem. In addition, deriving certified bounds that link embedding sensitivity, perturbation magnitude, attack success, and downstream performance degradation could provide deeper insights into these trade-offs and is an important direction for future work.

\bibliography{acml25} 

\newpage
\clearpage
\appendix

\section{Bounding Output Differences with LDP}\label{Appx.A}

Let $\mathcal{A}$ be a randomized mechanism, $R_{\theta}$ be the model parameter space, and $\epsilon$ denote the privacy budget associated with the noise scale  $\delta$. In $\delta$-\textsc{Steal}, we add LDP noise $Lap(0, \delta)$ to the token embeddings, which is equivalent to applying $\epsilon$-LDP to each data sample. By the post-processing property of LDP, the adversary's model $\theta_{adv}$ is also $\epsilon$-LDP. As a result, watermark detectors cannot determine whether $\theta_{adv}$ was trained on watermarked data and fail to verify its intellectual property, especially with small values of $\epsilon$. The difference between $\theta_{adv}$ trained with or without watermarked data is bounded as follows:
  $P[ \mathcal{A} (x+ Lap(0,\delta), y^{wm}+ Lap(0,\delta)) \in R_{\theta}]
   \nonumber \le   \exp^{\epsilon} P[ \mathcal{A} (x'+ Lap(0,\delta), y'^{wm}+ Lap(0,\delta)) = R_{\theta}]$.

\section{Privacy Budget Calculation}\label{Appx.B}

In $\delta$-\textsc{Steal}, we use a common LDP approach, which is a Laplace mechanism that adds Laplace noises into token embeddings of the model. The Laplace mechanism is defined as: $\mathcal{A}_{\mathcal{E}}(x, \mathcal{E}(x), \epsilon) = \mathcal{E}(x) + (L_1, L_2, \cdots, L_d) $,
where $\mathcal{E}(x)$ is the embedding of token $x$, $d$ is the embedding size, and $L_i$ are independent and identically distributed (i.i.d.) random variables drawn from Laplace noise centered at 0, scaled by $\delta = \frac{\Delta(\mathcal{E})}{\epsilon}$.   Given a noise scale $\delta$, to compute the privacy budget $\epsilon$, we first need to calculate the magnitude by which
a single individual's data can change the function $\mathcal{E}$ in the worst case, which is  defined as: $\Delta(\mathcal{E}) = \max_{  \forall  x, \tilde{x} \in N^{d}} \| \mathcal{E}(x) - \mathcal{E}(\tilde{x}) \|_1 $. 

For LLaMA-2, we obtain $\Delta(\mathcal{E}) = 0.3$, while for Mistral, $\Delta(\mathcal{E}) = 0.05$. With noise scales of $\delta \in {0.001, 0.01, 0.05, 0.1}$, the corresponding privacy budgets are:
\begin{itemize}
    \item[$\bullet$]  For LLaMA-2: $\epsilon = \frac{\Delta(\mathcal{E})}{\delta} = \{300, 30, 6, 3\}$
\item[$\bullet$] For Mistral: $\epsilon = \frac{\Delta(\mathcal{E})}{\delta} = \{50, 5, 1, 0.5\}$
\end{itemize}

\section{Supplemental Results}\label{Appx.C}
In this supplement, we include the results for the SemStamp watermark, which were not presented in the main body. SemStamp requires fine-tuning a robust sentence embedder to support the local sensitivity mechanism for sentence selection. Since we used the pre-trained embedder provided by the authors, the results are not optimal in our experiments.

\begin{figure}[t]
 \centering
\subfigure[AttackSR across LLMs]{\includegraphics[scale=0.3]{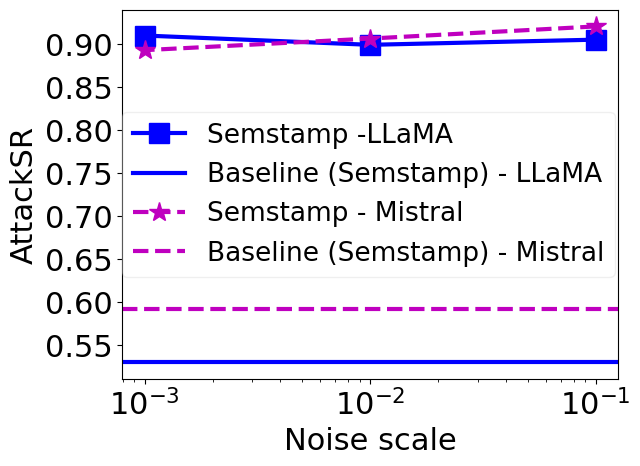}}\hspace{0.05cm} 
\subfigure[Perplexity (LLaMA-2)]{\includegraphics[scale=0.3]{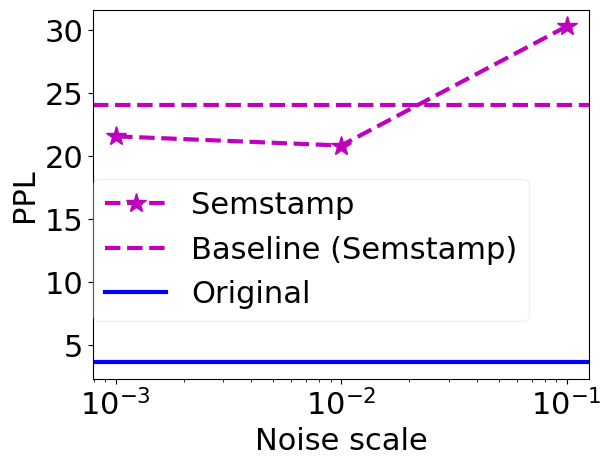}}
\subfigure[Perplexity (Mistral)]{\includegraphics[scale=0.3]{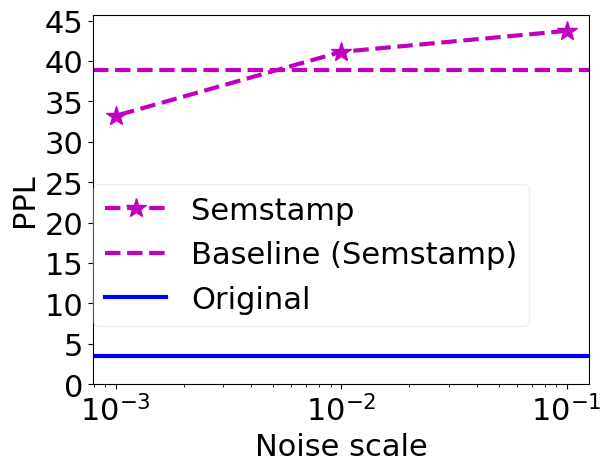}}
\hfill

\caption{AttackSR and Perplexity of $\delta$-\textsc{Steal} on Semstamp.} 
\label{fig:combined_watermark_evasion_sem}
\end{figure}

\noindent \textbf{$\delta$-\textsc{Steal} against Semstamp.}
For Semstamp, our observations of $\delta$-\textsc{Steal} attacks in Fig.~\ref{fig:combined_watermark_evasion_sem} are consistent with other watermarks. As noise scales increase, AttackSR rise, and PPL increases but remain comparable to the Baseline. However, a notable concern is the low Baseline watermark detection rate, which results in unexpectedly high AttackSR even without attacks, reaching $52.97\%$ for LLaMA-2 and $59.15\%$ for Mistral.

\begin{figure}[t]
\centering
\subfigure[LLaMA-2]{\includegraphics[scale=0.3]{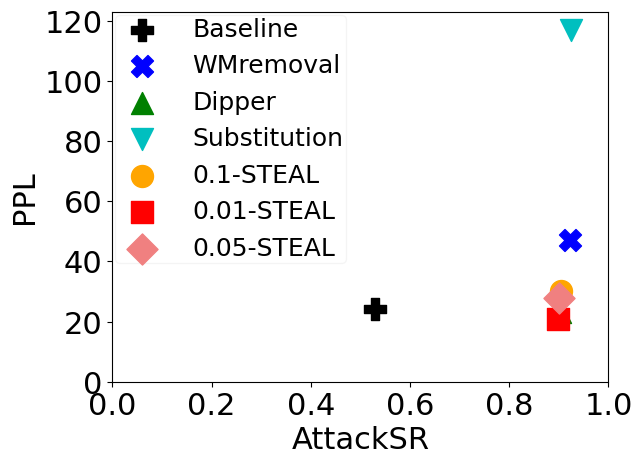}}
\hspace{0.05cm} 
\subfigure[Mistral]{\includegraphics[scale=0.3]{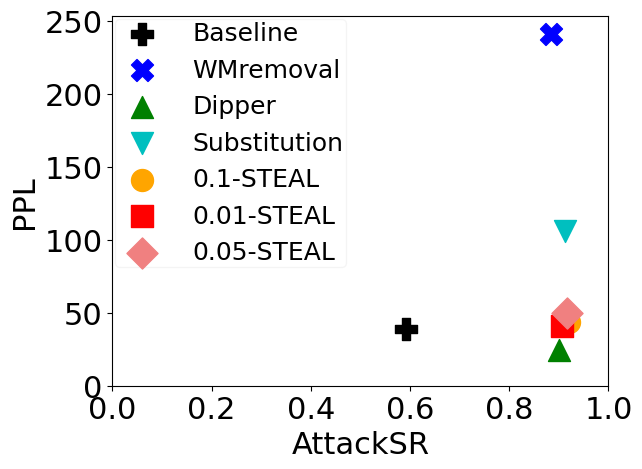}}
\hspace{0.05cm}

\caption{AttackSR and Perplexity across  on Semstamp.} 
\label{fig:compare_with_others_sem}
\end{figure}

\noindent \textbf{$\delta$-\textsc{Steal} and Existing   Attacks on Semstamp.}  
Fig.~\ref{fig:compare_with_others_sem} illustrates the effectiveness of our $\delta$-\textsc{Steal} attacks compared to other attacks on Semstamp. It is clear that $\delta$-\textsc{Steal}, regardless of noise scale, performs effectively, indicating high AttackSR and low PPLs.

\noindent  \textbf{$\delta$-\textsc{Steal} in MMLU Downstream Task.} 
In Table \ref{tab:model_utility_comparison}, we present only the KGW and EXP watermarking techniques, although we also experimented with all four techniques considered in this paper, including SIR and Semstamp. For SIR, our attempt to apply the publicly available trained model to the MMLU task resulted in poor accuracy, only 27.90\% for LLaMA-2 and 20.90\% for Mistral without any attacks. This poor performance is due to the fact that, due to  computational complexity, we do not retrain their watermark model, meaning it may not adapt well to our data and settings. For Semstamp, which generates entire sentences instead of individual tokens, it is not suitable for MMLU, which involves multiple-choice question-answering tasks.

\noindent  \textbf{Semantic Preservation of $\delta$-\textsc{Steal}  Outputs.}
We present additional quantitative examples from Mistral, as shown in Table 4a), to illustrate the effects of our $\delta$-\textsc{Steal} attack. These examples highlight the subtle differences introduced while preserving the overall semantic meaning. 
For instance, in the first row, the attack changes ``office has talked to Attorney General Jeff Sessions'' to ``team has talked with Jeff Sessions'' and removes the adverb ``now,'' while maintaining the key information about the Russia investigation. In the second example, ``20 percent of black students and 10 percent of Latino students in Boston are attending the city's top 20 public schools'' is modified to ``who applied to the city's top schools were admitted,'' subtly shifting the focus but retaining the core message that the mentioned students were accepted into top schools. Lastly, in the third instance, $\delta$-\textsc{Steal} changes ``shape of a new toy'' to ``shape, configuration, and appearance of a new product'' and modifies the quantifiers from plural to singular, changing ``patent'' and ``variety,'' while still discussing the types of patents. These examples demonstrate that our attack makes slight wording changes without altering the overall semantic content.

\newpage
Table 4b) provides a comparison of our $\delta$-\textsc{Steal} with the different attack methods used in this study. While the approach of our $\delta$-\textsc{Steal} attack has been detailed previously, Dipper and the Substitution attacks present alternative solutions. As per Dipper, since it works with sentence levels, it truncates and discards all incomplete sentences within the target output to attack and always starts paraphrasing full sentences only, resulting in a disrupted reading flow from input to output and causing small loss of information. For example, in the first row, while both $\delta$-\textsc{Steal} and Substitution attacks continue to generate ``JB'' to complete ``BJP'', Dipper starts with a new sentence. Furthermore, due to Dipper's reliance on reordering sentences, the provided snippet of its output within a limited token count shown in Table 4b could not fully show relevant benchmark against other attacks. For instance, in the second example, the sentence ``The company announced that it would pay $\$50,000$ for the silence of this alleged affair.'' introduces content beyond what is covered in the benchmark snippet.
In contrast, the Substitution attack replaces texts by considering surrounding context, maintaining a high degree of similarity (as shown by the larger green portion)  with the original watermarked text. 

\begin{table*}[t]
\centering
\scriptsize 
\captionsetup{labelformat=empty}
\caption{Table 4a): Examples of prompts and watermarked outputs with and without $\delta$-\textsc{Steal} (Mistral model). Green text highlights similarities and red text emphasizes differences.
}
\begin{tabular}{|p{0.18\linewidth}|p{0.30\linewidth}|p{0.30\linewidth}|p{0.04\linewidth}|p{0.035\linewidth}|}
\hline
\textbf{Prompt} & \textbf{Watermarked (W)} &  \textbf{Watermarked under $\delta$-Steal (SA) ($\delta = 0.01$) } & \textbf{PPL (W)} & \textbf{PPL (SA)} \\
\hline
\raggedright ... There have been a number of revelations this week related to the Russia investigation. Among them, that the special counsel's ... & 
\raggedright \textbf{{\color{blue} KGW:}}  \hlred{office} \hl{has talked} to \hl{attorney general Jeff Sessions} and that \hl{Robert Muller is} now \hl{looking} at \hlred{sitting down with} \hl{President} Donald \hl{Trump}. And, of course, \hlred{news about a deal to temporarily end} the \hl{government shutdown}, but with \hlred{no resolution} [...continues] & 
\raggedright \textbf{{\color{blue} KGW:}}  \hlred{team} \hl{has talked} with \hl{Attorney General Jeff Sessions}, and that \hl{Robert Mueller is} \hl{looking} to \hlred{sit down for an interview} with \hl{President} \hl{Trump}. But with a \hl{government shutdown} \hlred{still hanging over everyone's heads} , it can be easy to \hlred{lose track of} [...continues]  & 
8.52 & 8.21 \\
\hline
\raggedright ... Latino students as well saw less opportunity for access to quality schools than their white and Asian counterparts, ” O'Brien said. The study found that more than 80-... & 
\raggedright \textbf{{\color{blue} EXP:}}   \hl{percent of black students and 70-percent of Latino students} in \hlred{Boston} are attending schools that are \hlred{considered low performing}. \hl{The report also found that 20-percent of black students and 10-percent of Latino students} in \hlred{Boston} are \hlred{attending} the city' s \hl{top} 20 public \hl{schools} . `` The report [...continues] & 
\raggedright \textbf{{\color{blue} EXP:}} \hl{percent of black students and 70-percent of Latino students} \hlred{who applied} to \hlred{the city}' s top schools \hlred{were not admitted}. \hl{The report also found that 20-percent of black students and 10-percent of Latino students} who applied to \hlred{the city's} \hl{top} \hl{schools} were \hlred{admitted}. The report [...continues] 
  & 2.87 & 2.03 \\
\hline
\raggedright ...patent protects the functional aspects of an invention, such as a new machine. A design patent protects the ornamental appearance of an invention, such as the... & 
\raggedright  \textbf{{\color{blue} SIR:}} \hl{shape} of \hl{a new} \hlred{toy}. \hl{Plant patents protect}, \hlred{you guessed it}, \hl{new varieties of} plants. \hlred{You apply} for \hl{a patent} with \hl{the United States Patent} and \hlred{trademark Office}, but you can not put a patent symbol “ ® ” on your \hlred{product} until it is [...continues] & 
\raggedright  \textbf{{\color{blue} SIR:}}  \hl{shape}, configuration and appearance of \hl{a new} \hlred{product}. \hl{A plant patent protects} \hlred{inventions of natural organisms} such as a \hl{new variety of} fruit tree. Once you determine what type of \hlred{intellectual property} you need, \hlred{you can file} \hl{a patent} application with \hl{the U.S patent} [...continues] 
 & 5.67 & 5.52 \\
\hline
\end{tabular}%
\label{tab:3a}
\end{table*}
\setlength{\textfloatsep}{20pt}

\begin{table*}[t]
\centering
\scriptsize 
\captionsetup{labelformat=empty}
\caption{Table 4b): Examples of prompts and watermarked outputs of $\delta$-\textsc{Steal} and other attacks (LLaMA-2 model). Green text highlights similarities and red text emphasizes differences.}
\begin{tabular}{|p{0.14\linewidth}|p{0.19\linewidth}|p{0.18\linewidth}|p{0.17\linewidth}|p{0.18\linewidth}|}
\hline
\textbf{Prompt} &  \textbf{Watermarked (W)} &  \textbf{Watermarked under $\delta$-Steal (SA) ($\delta = 0.01$) } & \textbf{Dipper} & \textbf{Substitution} \\
\hline
\raggedright ... do that for the MCD polls, '' said a senior party leader. Another reason behind not targeting Modi in the MCD polls is the massive victory the B …  & 
\raggedright  \textbf{{\color{blue} KGW:}}  \hl{JP} \hlred{achieved} \hl{in the 2014 Lok Sabha} \hlred{polls} in \hlred{Delhi} , where the party won all seven seats. `` \hl{We have to} \hlred{take into account the ground reality} and the political equation in the city.  \hlred{We can not ignore} the \hl{BJP 's victory in the 2014 Lok Sabha} [...continues] & 
\raggedright  \textbf{{\color{blue} KGW:}} \hl{JP} \hlred{registered} \hl{in the 2014 Lok Sabha} \hlred{elections} in \hlred{the city}. `` \hl{We have to} \hlred{convince the people of Delhi} that why we are the best option to govern the city. \hlred{We have to show} that the \hl{BJP 's victory in the 2014 Lok Sabha} elections was a [...continues] & 
\raggedright   \textbf{{\color{blue} KGW:}} \hl{We have to} \hlred{study} \hl{the ground reality} and \hl{political} situation \hl{in the city}. \hlred{We can't} \hl{ignore} the sweeping \hl{victory} of the \hl{BJP} in the \hl{2014} elections \hl{in Delhi}. \hlred{The party won all the seven parliamentary seats.} [...continues] & 
 \textbf{{\color{blue} KGW:}} \hl{JP} \hlred{participated} \hl{in the 2014 Lok Sabha} elections \hl{in Delhi}, \hl{where the party won} just \hlred{two seats}. "We need to \hl{take into account} the political \hlred{climate} and the \hl{political} situation \hl{in the city}. \hl{We cannot} support the \hl{BJP's} \hlred{performance} \hl{in the 2014 Lok Sabha} \hlred{elections} [...continues]\\
\hline
\raggedright … Daniels, who was born Stephanie Clifford, was paid 130,000 by Cohen after she signed a nondisclosure agreement barring her from talking about her alleg... & 
\raggedright  \textbf{{\color{blue} EXP:}} \hl{ed affair with Trump}. Trump \hlred{denies the allegation}. \hl{Cohen} has admitted making \hl{the payment to Daniels}, which he said was done to protect \hlred{Trump' s campaign from the allegations}. It' s possible that the reimbursement \hl{payment} \hlred{revealed} in the \hl{financial disclosure} may have been [...continues] & 
\raggedright \textbf{{\color{blue} EXP:}}  \hl{ed affair with Trump}. \hl{Cohen} recently revealed that \hlred{Trump personally reimbursed him} for \hl{the payment to Daniels}. As a result of the payment to Cohen, the government ethics office \hlred{sent a reminder to Trump} that \hlred{he must disclose} in his annual \hl{financial disclosure} report [...continues] & 
\raggedright \textbf{{\color{blue} EXP:}}  \hlred{The company announced that it would pay \$ 50,000 for the silence of this} \hl{alleged affair}. \hl{Cohen admitted} \hlred{paying the money}, saying it was in order to \hl{protect Trump's campaign from} \hlred{the alleged affair}. \hl{Trump denies} \hlred{the alleged affair.} [...continues] &  \textbf{{\color{blue} EXP:}} 
\hlred{daniels spoke} \hl{with Trump}. \hl{trump denied the allegation}. \hl{Cohen} later \hl{admitted} to the \hl{payments to Daniels}, \hl{which he said were done to protect Trump's} \hlred{family} \hl{from the allegations}. \hl{It's possible that the} \hlred{cash} payments \hlred{mentioned} in the full \hl{disclosure} would [...continues] \\
\hline
\raggedright … other shows, it's a GoPro on a windshield, '' Foley said referring to Ride Along. `` I think if this was a show that was on... & 
\raggedright \textbf{{\color{blue} SIR:}}  \hlred{a traditional} \hl{television} platform, \hl{we would be able to do more with it}. As it stands, it 'll be a while before \hlred{we do any new episodes of Holy Folesy!} '' \hl{Foley also} \hlred{addressed} \hl{why his daughter Noelle is n't pursuing} a career in WWE despite [...continues] & 
\raggedright \textbf{{\color{blue} SIR:}}  \hlred{a major network} \hl{television}, \hl{we would be able to do more with it}. I think we would be able to have a bigger budget and \hlred{be able to do some cool things with it}. '' \hl{Foley also} \hlred{talked about} \hl{why his daughter Noelle is n't pursuing} an [...continues] & 
\raggedright \textbf{{\color{blue} SIR:}} 
 \hlred{Furthermore}, \hl{Foley} was asked \hl{why his daughter} \hlred{Nol} \hlred{was not continuing in the world of professional wrestling}, \hlred{despite her father's rich career}. \hlred{It's too early for her to become a wrestler, because the thing is that it's } [...continues] & 
\textbf{{\color{blue} SIR:}}  \hlred{a great wrestling} \hl{platform}, we would be happy \hl{to do} something \hl{with it}. \hl{as it stands}, \hl{it'll be a while before} we have any more fans of \hl{holy Folesy!"} \hl{Foley} \#\#a \hlred{explained} \hl{why his daughter Noelle isn't pursuing a career in WWE despite}. [...continues] \\
\hline
\end{tabular}%
\label{tab:3b}
\end{table*}

\newpage
\section{Ablation}\label{Appx.D}
To better understand the effect of noise injection location, we conduct a study and report results in Table~\ref{tab:ablation}  comparing three strategies on LLaMA-2 with KGW at $\delta=0.01$ on (1) noisy embeddings during fine-tuning ($\delta$-\textsc{Steal}), (2) noisy pre-logits, and (3) noisy embeddings applied at inference.

\begin{table}[h]
\small
\centering
\begin{tabular}{lcc}
\toprule
\textbf{Noise Location} & \textbf{PPL} & \textbf{AttackSR (\%)} \\
\midrule
Noisy Embedding (training, $\delta$-\textsc{Steal}) & 4.30 & 68.33\% \\
Noisy Pre-logits & 4.86 & 74.75\% \\
Noisy Embedding (inference) & 5.27 & 67.69\% \\
\bottomrule
\end{tabular}
\caption{Ablation study comparing noise injection locations on LLaMA-2 with KGW  at $\delta=0.01$.}
\label{tab:ablation}
\end{table}

\end{document}